\documentstyle{article}

\def\1ad{\mbox{\normalsize $^1$}}
\def\2ad{\mbox{\normalsize $^2$}}
\def\3ad{\mbox{\normalsize $^3$}}
\def\4ad{\mbox{\normalsize $^4$}}
\def\5ad{\mbox{\normalsize $^5$}}
\def\6ad{\mbox{\normalsize $^6$}}
\def\7ad{\mbox{\normalsize $^7$}}
\def\8ad{\mbox{\normalsize $^8$}}
\def\makefront{\vspace*{1cm}\begin{center}
\def\newtitleline{\\ \vskip 5pt}
{\Large\bf\titleline}\\
\vskip 1truecm
{\large\bf\authors}\\
\vskip 5truemm
\addresses
\end{center}
\vskip 1truecm
{\bf Abstract:}
\abstracttext
\vskip 1truecm}
\newcommand\be{\begin{equation}}
\newcommand\ee{\end{equation}} 
\newcommand\bea{\begin{eqnarray}}
\newcommand\eea{\end{eqnarray}}
\newcommand\ibar{{\bar\imath}}
\newcommand\jbar{{\bar\jmath}}
\newcommand\kbar{{\bar k}}

\newcommand\Vi{V^i}

\newcommand\Vibar{V^{\bar\imath}}
\newcommand\Vjbar{V^{\bar\jmath}}
\newcommand\Ri{P^i}
\newcommand\Rj{P^j}
\newcommand\Rk{P^k}
\newcommand\Fi{F^i}

\newcommand\Sibar{\Sigma^{\bar\imath}}
\newcommand\Sjbar{\Sigma^{\bar\jmath}}
\newcommand\Skbar{\Sigma^{\bar k}}
\newcommand\Dibar{\Delta^{\bar\imath}}
\newcommand\Djbar{\Delta^{\bar\jmath}}
\newcommand\Hibar{H^{\bar\imath}}
\newcommand\Hjbar{H^{\bar\jmath}}
\newcommand\GRi{i_{\gamma}(\Ri)}

\newcommand\GRk{i_{\gamma}(\Rk)}
\newcommand\s{{\cal S}}
\newcommand\Gammabar{{\bar\Gamma}}
\newcommand\MJ{{\cal M}_J}
\newcommand\Abar{\bar A}
\newcommand\alphabar{{\bar\alpha}}
\newcommand\half{{\textstyle{1\over2}}}
\begin {document}
\def\titleline{The Holomorphic Anomaly of Topological
Strings
}
\def\authors{
Carlo Becchi,  Stefano Giusto \1ad ,
Camillo Imbimbo \2ad
}
\def\addresses{
\1ad
Dipartimento di Fisica dell'Universit\`a di Genova \\
Via Dodecaneso 33, I-16146 Genova \\
\2ad
I.N.F.N. Sezione di Genova \\
Via Dodecaneso 33, I-16146 Genova \\
}
\def\abstracttext{
We show that the BRS operator of the topological string
B model is not holomorphic in the complex structure of the
target space. This implies that the so-called holomorphic anomaly
of topological strings should not be interpreted as
a BRS anomaly.
}
\makefront
\section{Introduction and Conclusions}

There exist two topological sigma models whose target space is 
a Calabi-Yau complex three-fold $X$ \cite{witten1}:
The A-model, which depends on the complexified K\"ahler structures
of $X$ and the B-model, which depends on its complex structures.
The moduli of both the A and the B model are in correspondence
with integrated two-forms which are closed under the
BRS operator characterizing the topological theory. 
The moduli spaces of both the A and the B model 
carry a natural complex structure; 
The operators corresponding to the holomorphic moduli are
non-trivial elements of the BRS cohomology, while those
associated to the anti-holomorphic ones are BRS-exact. 

This is understood in terms of the underlying 
$N=2$ superconformal model \cite{cv}: 
(Anti)holomorphic deformations correspond
to marginal operators 
$O^{(2)}_\alpha = {\tilde Q}{\bar{\tilde Q}}\, \Phi_\alpha$ 
($O^{(2)}_\alphabar = Q{\bar Q}\, \Phi_\alphabar$),
where $Q$ ${\tilde Q}$ are the two left-handed supersymmetry
charges, and ${\bar Q}$ ${\bar{\tilde Q}}$ are the right-handed
ones. $\Phi_\alpha$  ($\Phi_\alphabar$) are (anti)chiral
superfields. Upon twisting, $Q +{\bar Q} \equiv\, s$ is
identified with the BRS operator of the topological model: Thus from the
$N=2$ algebra one concludes that $O^{(2)}_\alpha$ is closed
but not exact, while the anti-holomorphic deformation $O^{(2)}_\alphabar
= s \, {\bar Q}\Phi_\alphabar $ is  BRS-trivial. 

Therefore, the derivatives with respect to the anti-holomorphic
moduli of generic physical correlators are equivalent to  BRS-trivial
insertions, and --- in absence of BRS anomalies --- should vanish.

Topological sigma models can be coupled to topological gravity to give
topological string models \cite{dvv}. In the conventional treatments, 
this coupling
looks relatively harmless: The BRS operator of the coupled system is
taken to be the sum of the BRS operator of the matter model with that
of the gravitational sector. The physical operators of the matter
system are therefore automatically promoted to elements of the
cohomology of the total $s$. Gravity couples to the matter via the 
term $\int \psi G$ in the action, where $\psi$ is the gravitino and
$G$ the matter supercurrent.  The world sheet super-reparametrization
symmetry is fixed by going to the superconformal gauge, after which
the gravitino $\psi$ becomes a differential on the moduli space of
Riemann surfaces ${\cal M}_g$ of given genus $g$: $\psi = d_m h$,
where $h$ is the two dimensional complex structure, $d_m$ is the
differential on ${\cal M}_g$. Thus the correlators $\langle
\prod_\alpha \int O^{(2)}_\alpha \rangle_{\rm coupled}$ of the coupled
system are reduced to integrals over ${\cal M}_g$ of the matter
correlators $\langle \bigl[\int \psi\, G]^{6g-6} \prod_\alpha 
\int O^{(2)}_\alpha \rangle_{\rm matter}$. In this framework the
insertion of BRS-trivial operators in the correlator produces
a total derivative on ${\cal M}_g$, since $\{ s, G \} = T$,
where $T$ is the stress-energy tensor, and the insertion of $\int \psi T$ is
equivalent to a derivative with respect to the Riemann moduli.
Thus it is expected that the derivative with respect to
an anti-holomorphic modulus of string physical correlators 
yields the integral over ${\cal M}_g$ of a total derivative. 
The computation of Cecotti et al. \cite{bcov}\ shows however that this integral
does not vanish. This has been therefore interpreted as the signal
that the boundary of ${\cal M}_g$ induces an anomaly of the BRS
symmetry.

This interpretation is problematic: The existence of a boundaryless
compactification of ${\cal M}_g$ is crucial to prove the very
consistency of string theory in general. It ensures independence
of physical correlators of gauge fixing  as well as of the choice
of the BRS representative for physical states. 

In fact we will show that, after 
coupling topological gravity to the topological B-model, 
the anti-holomorphic deformations become non-trivial elements of
the $s$-cohomology. 
This reconciles the anti-holomorphic dependence of topological
string correlators with the absence of any BRS anomaly.

Since the action of the B model is $s$-exact the complex structure 
dependence of the model is completely encoded in the BRS operator.
The $s$ of the rigid model is essentially the
Dolbeault operator $\bar\partial$ on $X$: 
This is indeed known to depend holomorphically
on the complex structure of $X$. However, as it will be shown in the 
following, the effect of topological gravity
is to deform $s$ by adding to it terms --- proportional to the
superghost $\gamma$ ---  which act as the anti-Dolbeault 
operator $\partial$ on $X$ \cite{bgi}. $\partial$ introduces an explicit
anti-holomorphic dependence in $s$. $\gamma$ is a field which vanishes
at the points where zero-form observables are inserted:
Therefore the cohomology of the deformed $s$ on the local
operators is unchanged and matter observables do remain physical.
However when $s$ acts on integrated operators like the action
the superghost does not vanish. Thus the anti-holomorphic derivatives
of the action pick up a term which is not $s$-trivial.

One can indeed work in a gauge --- like the superconformal one ---
which sets $\gamma$ to zero. With this choice antiholomorphic derivatives
of the action do appear to be $s$-trivial. However in this gauge
correlators are not globally defined forms on moduli space and contact
terms --- which introduce the antiholomorphic dependence on the $X$ complex
structure  --- are required to restore gauge invariance \cite{bi}. 
In conclusion, the $s$-triviality of the antiholomorphic derivatives is a 
mere gauge artifact.  
 
\section{Coupling the B-model to topological gravity}

Let $X$ be a Calabi-Yau complex three-fold, and $J$ a complex
structure on  it defined by a system of complex coordinates 
$(\Vi, \Vibar)$. 
The topological rigid B-model on $X$   is characterized by the 
following BRS transformations \cite{witten1}:
\be
\begin{array}{ll}
s\, \Vi =  0      &\qquad\qquad s\, \Vibar = \Sibar \\
s\, \Ri = d\, \Vi &\qquad\qquad s\, \Sibar  =  0 \qquad\qquad  s\, \Dibar = \Hibar \\
s\, \Fi = d\, \Ri &\qquad\qquad s\, \Hibar =  0 
\label{rigidbrs}
\end{array}
\ee
$\Ri \equiv P_\mu^i dx^\mu$ is a fermionic one-form on the world sheet
$\Sigma$.
$\Fi \equiv {1\over2}\Fi_{\mu\nu}dx^\mu dx^\nu$ is a bosonic auxiliary
field with values in the two-forms on the world sheet. 
$\Sibar , \Dibar$ are fermionic scalar fields. $\Hibar$ is
a bosonic scalar auxiliary field. $\Vi$ and $\Vibar$ are fields of
ghost number zero: The ghost number assignment of the other fields in 
(\ref{rigidbrs}) is fixed by the fact that the BRS operator $s$
carries ghost number 1.

The action $S$ of the B-model is s-trivial, $S = s W$. The standard
choice for the gauge fermion $W$ is :
\be
W = \int_\Sigma G_{i\jbar}\bigl( \Fi \Djbar + \star \Ri d\,\Vjbar\bigr),  
\label{gaugefermion}
\ee
where $G_{i\jbar}$ is a (Ricci-flat) K\"ahler metric on $X$ adapted
to the complex structure $J$. To keep invariance under holomorphic
reparametrizations of $X$ explicit at each stage of the computations,
it is convenient to modify the BRS operator $s$ in (\ref{rigidbrs})
as follows
\be
\begin{array}{ll}
s\, \Vi = 0 &\quad s\, \Vibar = \Sibar \\
s\, \Ri = d\, \Vi &\quad s\, \Sibar = 0\qquad\qquad\quad s\, \Dibar = \Hibar - 
\Skbar \Gammabar^{\ibar}_{\kbar \jbar}\Djbar \\
s\, \Fi = D\, \Ri  + \half \Sibar\Rk R^i_{\ibar k j}\Rj &\quad
s\, \Hibar = -\Skbar \Gammabar^{\ibar}_{\kbar \jbar}\Hjbar ,
\label{rigidcovariantbrs}
\end{array}
\ee
where $\Gamma^i_{jk}$ and $\Gammabar^{\ibar}_{\jbar\kbar}$ are the components
of the Levi-Civita connection relative to $G_{i\jbar}$. 
It is then natural to define the {\it covariant}  operator 
$\s \equiv s + \Sibar \Gammabar_{\ibar}$,  where the connection term
is understood to act on fields --- like $\Dibar$ and $\Hibar$ --- with 
values in the antiholomorphic tangent of $X$. 
$\s$ is nilpotent because the curvature two-form associated to
$\Gamma$ is of type $(1,1)$.

The topological rigid B-model can be coupled to topological
gravity, whose fields are the world sheet metric $g_{\mu\nu}$,
the gravitino $\psi_{\mu\nu}$, the diffeomorphism ghost $c^\mu$
and the commuting superghost $\gamma^\mu$ of ghost number 2.
Their BRS transformation laws are \cite{bi}:
\bea
s\, g_{\mu\nu} &=& \psi_{\mu\nu}-{\cal L}_c g_{\mu\nu} \qquad\quad \, 
s\, c^{\mu} \, = \, \gamma^{\mu} -\half{\cal L}_c c^{\mu}\nonumber \\ 
s\, \psi_{\mu\nu} &=& {\cal L}_{\gamma} g_{\mu\nu}-{\cal L}_c \psi_{\mu\nu}
\qquad
s\, \gamma^{\mu} \, = \, - {\cal L}_c\gamma^{\mu},
\label{topgravitybrs}
\eea
where ${\cal L}_c$ denotes the Lie derivative with respect to the vector
field $c^\mu$. After coupling to topological gravity, the BRS transformation
laws of the matter fields must include an infinitesimal diffeomorphisms with
parameter $c$.  Since $c$ transforms into $\gamma$, extra terms are required
to ensure the nilpotency of $s$. Thus, for example, the BRS transformation
rules of $\Vi$ become
\be
s \, \Vi = - {\cal L}_c \Vi + \cdots =  - i_c d\Vi +\cdots,
\ee
where $i_c$ is the contraction of a form with the vector field
$c$. Hence,
\be
s^2 \, \Vi = 0 = -{\cal L}_\gamma \Vi + s\, (\cdots),
\ee
from which one concludes that
\be
s \, \Vi =  - {\cal L}_c \Vi + \GRi ,
\ee
given that $ s\, \Ri = d \Vi +\cdots$. Proceeding in this way one
arrives to the BRS rules for the coupled system:
\bea
&& \s\, \Vi =  \GRi \nonumber\\
&& \s\, \Ri = d\, \Vi + i_{\gamma}(\Fi) \nonumber\\
&& \s\, \Fi = D\, \Ri  + \half \Sibar\Rk R^i_{\ibar k j}\Rj \nonumber\\
&&\\
&& \s\, \Vibar = \Sibar \nonumber\\
&& \s\, \Sibar = {\cal L}_{\gamma}\Vibar \nonumber \\
&& \s\, \Dibar = \Hibar \nonumber \\
&& \s\, \Hibar = i_\gamma (D\,\Dibar) + \GRi\Skbar 
{\bar R}^\ibar_{i\kbar \jbar}\Djbar
\label{coupledbrs}
\eea
To simplify notations, we have modified the definition of the
covariant operator $\s \equiv s + \Sibar \Gammabar_{\ibar} +
\GRi \Gamma_i + {\cal L}_c$, to include both the Levi-Civita
connection acting on fields with values in the holomorphic
tangent of $X$ and the action of world sheet infinitesimal diffeomorphisms
with parameter $c$. The nilpotency of the $s$ for the coupled
system implies that the square of the covariant $\s$ is:
\be
\s ^2 = \{ D, i_\gamma \} + \GRi \Sjbar R_{i\jbar}
\label{curvature}
\ee
The r.h.s. of (\ref{curvature}) is the curvature of the infinite-dimensional
bundle in field space which underlies the geometry of the system:
The total space of this bundle is the product of the space of complex
structures of Riemann surfaces $\Sigma$ and the tangent space to
the space of maps from $\Sigma$ into $X$.
Its structure group is the product of the group of diffeomorphisms of $\Sigma$
and the group of holomorphic reparametrizations of $X$.

\section{Dependence of $s$ on the complex structure of $X$}

The action of the B-model is $s$-trivial. Therefore the  physical
correlators are independent of the choice of the K\"ahler metric 
$G_{i\jbar}$ which appears in Eq. (\ref{gaugefermion}). However
the definition of the BRS operator $s$ requires choosing a
complex structure $J$ on $X$: Physical correlators will
be therefore functions of $J$. 

Because of the triviality of the action, the functional measure
for physical correlators  localizes on the field zero modes.
In particular, when discussing operators depending on world sheet
scalars --- such as $\Vi$, $\Vibar$, $\Sibar$, $\Dibar$ --- 
we can assume that the fields are constant.

Before coupling to topological gravity, $s$ acts on functions of
$\Vi, \Vibar , \Sibar$ as the Dolbeault differential, since
\be
s\, O(\Vi, \Vibar , \Sibar) = \Sibar\partial_{\Vibar} O(\Vi, \Vibar, \Sibar),
\label{dolbeault}
\ee
Thus, to each element $\omega_{{\ibar}_1\ldots{\ibar}_q}(V,{\bar V})
\, dV^{\ibar_1}\ldots dV^{\ibar_q}$
of the  $H^{(0,q)}(X)$ Dolbeault cohomology of 
forms on $X$ of type $(0, q)$, there 
corresponds an observable $\omega_{{\ibar}_1\ldots{\ibar}_q}(V,{\bar V})
\, \Sigma^{\ibar_1}\ldots \Sigma^{\ibar_q}$ of ghost number $q$.

Moreover 
\be
s\, (G_{i\jbar}\Djbar) = G_{i\jbar}\Hjbar \equiv H_i
\label{auxiliary}
\ee
$H_i$ appears in the action as an auxiliary field and the equation of
motion relative to $\Fi$ gives: $H_i = G_{i\jbar}\, i_\gamma 
(\star d\Vjbar)$. Thus, as long as one
considers {\it local} operators which do not involve $\Fi$, it is legitimate
to take $s \, \Delta_i = 0$, with $\Delta_i\equiv G_{i\jbar}\Djbar$.
It follows that $\omega^{i_1\ldots i_p}_{{\jbar}_1\ldots{\jbar}_q}(V,{\bar V})
\,\Delta_{i_1}\ldots \Delta_{i_p} \Sigma^{\ibar_1}\ldots \Sigma^{\ibar_q}$
is  an element of the cohomology of $s$ if 
$\omega^{i_1\ldots i_p}_{{\jbar}_1\ldots{\jbar}_q}(V,{\bar V})
\,\partial_{V^{i_1}}\ldots \partial_{V^{i_p}} dV^{\ibar_1}\ldots dV^{\ibar_q}$
is an element of the cohomology $H^{(0,q)}(X,\Lambda^p T^{(1,0)}X)$ of
$(0,q)$ forms with values in the bundle $\Lambda^p T^{(1,0)}$,
the $p^{\rm th}$ exterior power of the holomorphic tangent bundle $T^{(1,0)}
X$.

Among the observables of this class, those relative to elements of 
$H^{(0,1)}(X,T^{(1,0)}X)$  correspond to holomorphic 
deformations of the complex structure of $X$. Let us choose a basis for them
\be
O_\alpha = \Delta_i \mu^i_{\alpha\jbar}\Sjbar \equiv \Delta^{\rm t}
\mu_{\alpha}{\bar\Sigma}
\label{observables}
\ee
where $\alpha =1,\ldots, {\rm dim} H^{(0,1)}(X,T^{(1,0)}X)$ and we
introduced an obvious matrix notation for $\mu^i_{\alpha\jbar}$,
$\Delta_i$ and $\Sjbar$ and ${\rm t}$ denotes the transpose.

An explicit construction of $\mu^i_{\alpha\jbar}$ makes use of the
Beltrami  parametrization for $J$ given by the equations:
\be
d\, V = \Lambda (d\, v  + \mu d\,{\bar v})\qquad d\, {\bar V} = 
{\bar\Lambda} (d\, {\bar v}  + {\bar\mu} d\,v) 
\label{beltrami}
\ee
where $(v, {\bar v}) \equiv (v^i, v^\ibar)$ is a fixed system of complex 
coordinates and $\mu \equiv \mu^i_\jbar$, ${\bar\mu}\equiv
{\bar\mu}^\ibar_j$ are the Beltrami differentials.
A natural complex structure on  $\MJ$, the moduli space of complex structures
on $X$, is defined by declaring $\mu $ (${\bar\mu}$)
to be a (anti)holomorphic function of $J$. Let $(m_\alpha , m_{\bar\alpha})$
be complex coordinates on $\MJ$ . The holomorphic derivative
of $d\,V$ with respect to $m_\alpha$ is a form of mixed type \cite{cd}:
\be
\partial_\alpha d\, V = A_{\alpha}d\,V + \mu_\alpha d\, {\bar V} 
\label{kodaira}
\ee
Since $\mu_\alpha d\, {\bar V} = {\bar\partial}\partial_\alpha V$,
it follows that $\mu_\alpha d\, {\bar V}$ is indeed an element of
$H^{(0,1)}(X,T^{(1,0)}X)$.  By plugging Eq. (\ref{beltrami})
into Eq. (\ref{kodaira}), one can write $A_\alpha$ and $\mu_\alpha$
in terms of the Beltrami differentials:
\bea
A_\alpha &=& \partial_\alpha \bigl(\Lambda A^{-1}\bigr) A\Lambda^{-1} 
\nonumber\\
\mu_\alpha &=& \Lambda \partial_\alpha\mu \Abar {\bar\Lambda}^{-1},
\eea
where $A \equiv \bigl( 1 - \mu \,{\bar\mu}\bigr)^{-1}$.

Let us use the Beltrami parametrization to exhibit the dependence of 
the Dolbeault operator ${\bar\partial}_J \equiv d\Vibar \partial_{\Vibar}$
on $J$:
\be
{\bar\partial}_J = (d\, {\bar v} + {\bar\mu} d\, v)^{\rm t} 
\Abar^{\rm t}(\partial_{\bar v} - \mu^{\rm t}\partial_v)
\equiv {\rm {\bar e}}^{\rm t}(\partial_{\bar v} - \mu^{\rm t}\partial_v)
\ee
where  ${\rm e}^\ibar =
[\Abar (d\, {\bar v} + {\bar\mu} d\, v)]^\ibar$ is a basis for the one-forms
of type $(0,1)$ in the complex structure $J$.
This redefinition of the basis of one-forms of type $(0,1)$ 
renders ${\bar\partial}_J$ manifestly holomorphic in $J$.

Thus, working in this basis, the elements of the Dolbeault cohomology
--- i.e. the solutions of ${\bar\partial}_J \omega = 0$ ---
can be taken to depend holomorphically on $J$. This means that
the vector bundle over $\MJ$ whose fiber is $H^{(0,q)}(X)$
is a holomorphic vector bundle. The same is true for the
vector bundle with fiber $H^{(0,q)}(X,\Lambda^p T^{(1,0)}X)$.

This implies that the BRS operator for the B-model before coupling
to topological gravity does not depend on $\bar\mu$, as long as one
makes the redefinition
\be
{\overline\Sigma} = {\bar\Lambda}\Abar^{-1} {\bar\sigma} 
\ee
in the anti-holomorphic sector. Furthermore the action
of the BRS operator on the new fields should not depend on the non-local
integrating factor $\Lambda$. It is easy to verify that this dictates
the following redefinitions in the holomorphic sector: 
\be
\Delta^{\rm t} = \delta^{\rm t}\Lambda^{-1}\quad H^{\rm t} = h^{\rm t}
\Lambda^{-1}\quad P = \Lambda \rho \quad F = \Lambda f
\ee
In this field basis, the observables (\ref{observables}) become:
\be
O_\alpha = \delta^{\rm t}\partial_\alpha \mu {\bar\sigma},
\ee
which shows explicitely that in the new field basis one can
choose a set of observables which are holomorphic in $J$.

After coupling to topological gravity the action of $s$ is
modified to be:
\bea
s &=& \Sibar\partial_{\Vibar} + \GRi \partial_{\Vi} + \ldots \nonumber\\
  &=& {\bar\sigma}^{\rm t}\bigl(\partial_{\bar v} - \mu^{\rm t}
\partial_v\bigr) + i_\gamma(A\rho)^{\rm t}(\partial_v - {\bar\mu}^{\rm t}
\partial_{\bar v}\bigr) +\ldots , 
\label{deformed}
\eea
and $\s \,{\bar\Sigma} = {\cal L}_\gamma {\bar V}$,  
$\s \, i_\gamma(P) = {\cal L}_\gamma V$. 

The ghost $c$ and the superghost $\gamma$ vanish on the points of
the world sheet where local zero-forms  observables are inserted.  Thus
$s$ reduces on such observables to the $s$ of the rigid theory.
In particular, the operators (\ref{observables}) remain physical
after coupling to topological gravity. 

However, on operators integrated over the world sheet, such as
the action, the $\gamma$ dependent term in Eq. (\ref{deformed}) 
does not vanish: It induces a dependence of $s$ on 
$\bar\mu$. This is the origin of the holomorphic anomaly of the
topological string B model.  

Let us evaluate the dependence of $s$ on $\bar\mu$:
\be
I_{\alphabar} \equiv  \bigl[\partial_{\alphabar} , \s \bigr] = 
\GRi \bigl[\partial_{\alphabar} , \partial_{\Vi} + \ldots \bigr] 
= - \GRi \mu_{\alphabar i}^\jbar \partial_{\Vjbar} +\ldots 
\ee
This gives for the antiholomorphic dependence of the action the following
result:
\be
\partial_\alphabar\, S = I_\alphabar\, W + s \bigl(\partial_\alphabar\, W
\bigr) = \int_\Sigma G_{i\jbar}\, D\!\star\!\Ri\, \mu_{\alphabar k}^\jbar\,
\GRk
+ s \bigl(\partial_\alphabar\, W \bigr),
\ee  
from which one derives the Ward identity:
\be
\partial_\alphabar \langle \prod_\alpha O_\alpha \rangle =
\langle \int_\Sigma G_{i\jbar}\, D\!\star\!\Ri\, \mu_{\alphabar k}^\jbar\,
\GRk \prod_\alpha O_\alpha \rangle .
\label{ward}
\ee
Let us end by remarking that $I_\alphabar\, W$ is BRS-closed modulo
the equation of motions since
\be
s\, I_\alphabar\, W = - I_\alphabar s\, W = - I_\alphabar\, S,  
\ee
which ensures that the correlator in the r.h.s. of Eq. (\ref{ward})
is gauge-invariant.



\begin{thebibliography}{77}
\bibitem{witten1} E. Witten, ``Mirror Manifolds and Topological Field
Theory'',  in {\it Essays on Mirror Manifolds}, edited by S. Yau,
International Press (Honk Kong 1992).

\bibitem{cv} 
S. Cecotti and C. Vafa, {\it Nucl. Phys.} {\bf B367} (1991) 359. 

\bibitem{dvv}
R. Dijkgraaf, E. Verlinde and H. Verlinde, ``Notes on Topological
String Theory and 2-D Quantum Gravity'', in {\it String Theory and 
Quantum Gravity `90}, proceedings of the 
Trieste Spring School and Workshop, ICTP, Trieste, 1990, World Scientific,
(Singapore 1991). \\
E. Witten, {\it Nucl. Phys} {\bf B340} (1990) 281. 

\bibitem{bcov} M. Bershadsky, S. Cecotti, H. Ooguri and C. Vafa,
{\it Commun. Math. Phys.} {\bf 165} (1994) 311.

\bibitem{bgi} 
C.M. Becchi, S. Giusto and C. Imbimbo,
``The BRST structure of twisted $N=2$ algebra'', forthcoming in the 
proceedings of the 
{\it International Conference on Secondary Calculus and Cohomological Physics},
Moscow, 24-31 August 1997, Contemporary Mathematics series of
the American Mathematical Society.

\bibitem{bi}
C.M. Becchi and C. Imbimbo, {\it Nucl. Phys.} {\bf B462} (1996) 571.

\bibitem{cd}
P. Candelas and X. de la Ossa, {\it  Nucl.Phys.} {\bf B355} (1991) 455.
\end{thebibliography}
\end{document}